\documentclass [12pt,preprint]{aastex}
\usepackage{epsfig}
\begin{document}
\voffset-0.5cm
\newcommand{\gsim}{\hbox{\rlap{$^>$}$_\sim$}}
\newcommand{\lsim}{\hbox{\rlap{$^<$}$_\sim$}}

\title {High Energy Photons From Gamma Ray Bursts}

\author{Shlomo Dado\altaffilmark{1} and Arnon Dar\altaffilmark{2}}

\altaffiltext{1}{dado@phep3.technion.ac.il\\
Physics Department, Technion, Haifa 32000,
Israel}
\altaffiltext{2}{arnon@physics.technion.ac.il\\
Physics Department, Technion, Haifa 32000}

\begin{abstract} 

Emission of high energy (HE) photons above 100 MeV that is delayed and 
lasts much longer than the prompt MeV emission has been detected from 
several long duration gamma ray bursts (LGRBs) and short hard bursts 
(SHBs) by the Compton, Fermi and AGILE gamma ray observatories. In this 
paper we show that the main observed properties of this HE emission are 
those predicted by the cannonball (CB) model of GRBs: In the CB model all 
the observed radiations in a GRB are produced by the interaction of a 
highly relativistic jet of plasmoids (CBs) with the environment. The 
prompt X-ray and MeV $\gamma$-ray pulses are produced by inverse Compton 
scattering (ICS) of glory photons -photons scattered/emitted into a cavity 
created by the wind/ejecta blown from the progenitor star or a companion 
star long before the GRB- by the thermal electrons in the CBs. A 
simultaneous optical and high energy emission begins shortly after each 
MeV pulse when the CB collides with the wind/ejecta, and continues during 
the deceleration of the CB in the interstellar medium. The optical 
emission is dominated by synchrotron radiation (SR) from the swept-in and 
knocked-on electrons which are Fermi accelerated to high energies by the 
turbulent magnetic fields in the CBs, while ICS of these SR photons 
dominates the emission of HE photons. The lightcurves of the optical and 
HE emissions have approximately the same temporal behaviour but have 
different power-law spectra. The emission of very high energy (VHE) 
photons above 100 TeV is dominated by the decay of $\pi^0$'s produced in 
hadronic collisions of Fermi accelerated protons in the CBs. The CB model 
explains well all the observed radiations, including the high energy 
radiation from both LGRBs and SHBs as demonstrated here for GRB 090902B 
and SHB 090510.

\end{abstract}

\maketitle

\section{Introduction}

During nearly 20 years after the launch of the Compton Gamma Ray 
Observatory (CGRO), the Burst And Transient Source Experiment (BATSE) on 
board the CGRO has detected and measured light curves and spectra (Kaneko 
et al.~2008)  in the sub-MeV range of several thousands gamma ray bursts 
(GRBs). Higher-energy observations with the EGRET instrument aboard CGRO 
were limited to those GRBs which happened to be in its narrower field of 
view. Its large calorimeter measured the light-curves and spectra of 
several GRBs in the 1-200 MeV energy range. Seven GRBs were detected also 
with the EGRET spark chamber, sensitive in the 30 MeV - 10 GeV energy 
range. The EGRET detections indicated that the spectrum of bright GRBs 
extends beyond 1 GeV (Hurley et al.~1994) with no evidence for a spectral 
cut-off (see, e.g., Dingus~1995,~2001 and references therein). However, a 
few GRBs, such as 940217 (Hurley et al.~1994) and 941017 (Gonzalez et 
al.~2003), showed evidence that the high energy component has a slower 
temporal decay than that of the sub-MeV emission, suggesting that, at 
least in some cases, it is not a simple extension of the main component, 
but originates from a different emission mechanism and/or region. This has 
been confirmed recently by observations of high energy photons in the 30 
MeV - 300 GeV range from several GRBs with the AGILE (GRB 080514B: 
Giuliani et al.~2008, GRB 090510: Giuliani et al.~2009) and the Fermi 
large area telescope (LAT) (e.g., GRB 080916C: Abdo et al.~2009a, GRB 
090902B: Bissaldi et al.~2009 and GRB 090510: Ghirlanda et al.~2009). The 
arrival times of the high energy photons did not coincide with the times 
of the brightest peaks seen at hard X-rays and MeV $\gamma$-rays. Also the 
high energy emission lasted much longer time than that of the prompt 
keV-MeV emission.

The detection of higher energy gamma rays is affected by pair production 
in their collisions with the extragalactic infrared background light 
(Nikishov 1961) resulting in an absorption which is a strong function of 
redshift and energy. Recent estimates (Primack et al.~2005), validated by 
HESS observations (Aharonian et al.~2006) predict an optical depth of 
roughly unity to 500 GeV photons emitted at a redshift $z\!=\!0.2$ and to 
10 TeV photons at $z=0.05$. The average redshifts of LGRBs and SHBs are 
much larger, $z\!=\!2.2$ and $z\!=\!0.5$, respectively. Despite of the 
strong attenuation of high energy gamma rays in the intergalactic medium 
(IGM), there have been several claims in the past of detections at the 3 
sigma level of TeV gamma-rays from GRBs (see, e.g., Atkins et al.~2003 and 
references therein). However, more recently, no GRB was conclusively 
detected in the range 100 GeV to 100 TeV by the ground based water 
Cherenkov detector Milagro and by the air Cherenkov telescopes MAGIC, 
Whipple, HESS and VERITAS. Moreover in all previous cases of reported 
detection of TeV gamma-rays from a GRB, the GRB redshift was not known. If 
their redshifts are similar to those of ordinary GRBs then their TeV 
gamma-rays are strongly absorbed by the extragalactic background light 
(EBL), implying that the TeV emission if detected must be extraordinarily 
energetic, i.e., with a much larger fluence than that emitted in X-rays 
and MeV $\gamma$-rays.

Most theoretical models of high energy photon emission  
in GRBs relied on the standard fireball 
models of GRBs (for recent reviews see, e.g., Piran~2005; 
M\'{e}sz\'{a}ros~2006; Zhang~2007). In these models, synchrotron radiation 
of electrons accelerated by `internal shocks' in collisions between 
conical shells ejected by the GRB's central engine produces the prompt 
GRB emission and a blast wave (external shock) driven into the circumburst 
medium generates their afterglow.  The high energy radiation was suggested 
to be produced either by inverse Compton scattering of the synchrotron 
radiation (the so called `synchrotron self Compton mechanism') by the 
shock-accelerated electrons (e.g., Dermer et al.~2000), by the decay of 
$\pi^0$ photo produced in collisions of shock accelerated hadrons with 
synchrotron photons in the expanding fireball (Waxman \& Bahcall~1997), or 
by synchrotron radiation from ultra high energy protons (Totani~1998) 
allegedly accelerated in the GRB fireball (Waxman~1995; Milgrom \& 
Usov~1995; Vietri~1995). All these models that were based on the standard 
fireball model of GRBs predict simultaneous emissions at all energy bands. 
This is in conflict with the observed delayed emission of the high energy 
photons that lasts much longer. 

In the cannonball (CB) model (see e.g. Dado, Dar \& De R\'ujula, hereafter 
DDD, 2009a,b and references therein), GRBs and their afterglows (AGs) are 
produced by bipolar jets of highly relativistic plasmoids (CBs) ejected in 
violent stellar processes. The prompt MeV $\gamma$-rays and hard X-rays 
are produced by inverse Compton scattering (ICS) of glory photons - 
photons emitted/scattered into a cavity formed by the wind/ejecta puffed 
by the progenitor or a companion star long before the GRB. Synchrotron 
radiation (SR) emitted from the electrons of the ionized wind/ejecta that 
are swept into the CBs and are Fermi accelerated by their turbulent 
magnetic fields dominates their `prompt' optical emission (e.g., DDD2009a 
and references therein) which begins when the CBs reach the wind/ejecta. 
In this paper we show that in the CB model ICS of this SR (e.g., Dado \& 
Dar~2005) dominates the HE emission from long GRBs and SHBs. This HE 
emission begins simultaneously with the `prompt' optical emission that 
lags after the prompt X-ray and MeV $\gamma$-ray emission and lasts much 
longer, has the same lightcurve as the optical emission but with a 
power-law spectrum that is identical to that of the X-ray emission. It 
describes well the HE observations, as demonstrated in the paper for GRB 
090902B and SHB 090510.

Because of the Klein-Nishina effect, $E^2\,dn/dE$ of ICS of the prompt 
SR in GRBs peaks near 10 TeV and is cutoff at a few tens of TeV when the 
rate of synchrotron energy losses by electrons in the CBs exceeds the rate 
of energy gain by Fermi acceleration.  The prompt decay of $\pi^0$'s 
produced in the collisions between the Fermi accelerated nuclei and the 
ambient matter in the CBs produces a power-law spectrum that extends to 
much higher energies where it dominates the HE emission. However, like in 
blazars, the observed flux of TeV photons from distant GRBs and SHBs is 
stronglyly suppressed by pair production in collisions with the 
extragalactic background photons and only relatively nearby GRBs and SHBs 
might be detected in TeV photons by the large ground based HE gamma ray 
telescopes such as HESS, MAGIC and VERITAS.

\section{The CB model} 

\begin{figure}[]
\centering
\epsfig{file=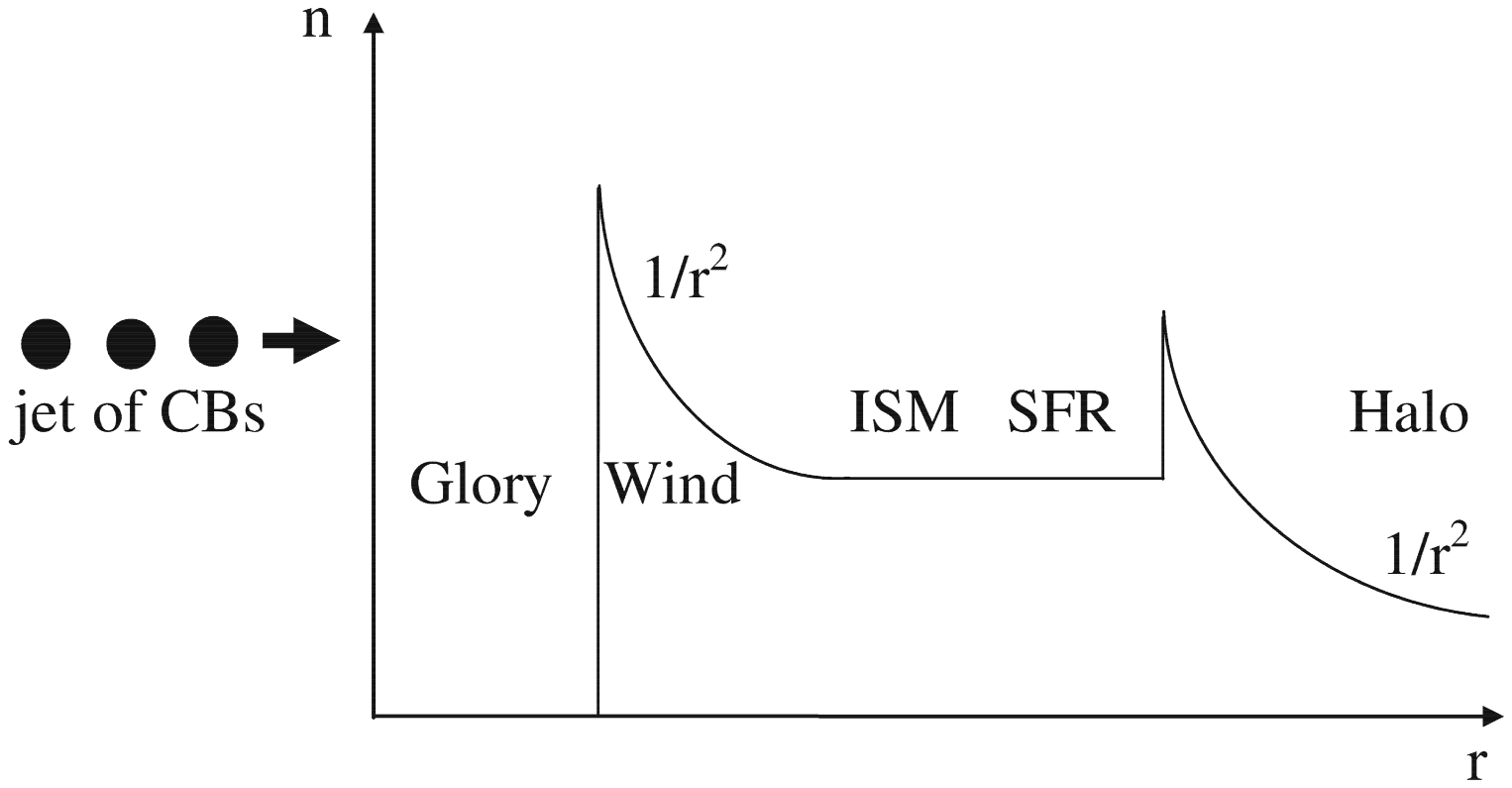,width=18cm}
\caption{Schematic illustration, not in scale, of the typical environment 
encountered by 
a highly relativistic jet ejected in core collapse SN that escapes 
from the star formation region into the galactic halo.}    

\label{f0}
\end{figure}

In the cannonball (CB) model (Dado, Dar \& De R\'ujula (hereafter DDD) 
2002; Dar \& De R\'ujula (hereafter DD) 2004; DDD2009a,b and references 
therein) GRBs and their AGs are produced by bipolar jets of highly 
relativistic CBs of ordinary matter which are ejected (Shaviv \& Dar~1995, 
Dar \& Plaga~1999) in the birth of neutron stars or black holes in 
core-collapse supernova (SN) explosions (long GRBs) akin to SN 1998bw 
(Galama et al.~1998), and in the merger of neutron stars and/or phase 
transition in compact stars (short hard bursts). Their prompt MeV 
$\gamma$-rays and hard X-rays are produced by the thermal electrons in the 
CBs' plasma via inverse Compton scattering (ICS) of glory photons - 
photons emitted/scattered into a cavity created by the wind/ejecta blown 
from the progenitor star or a companion star long before the GRB. Slightly 
later when the CBs encounter the wind/ejecta, and afterwards when the CBs 
coast through the interstellar medium (ISM) sorounding it, the electrons 
of the ionized gas in front of them that are swept in and Fermi 
accelerated by the CBs' turbulent magnetic fields emit synchrotron 
radiation (SR) which dominates the `prompt' optical emission and the broad 
band afterglow emission. ICS of the SR radiation by these electrons and 
the decay of $\pi^0$'s produced in collision between the swept-in wind and 
ISM protons and the ambient CB protons produce the `prompt' high energy 
emission simultaneously with the optical emission.  Within the CB model, 
the burst environment as illustrated in Fig.~(\ref{f0}) and the above 
radiation mechanisms, which are summarized in Table~\ref{t1}, suffice to 
provide a sufficiently accurate description of the observed radiations 
from GRBs at all times and all detected wavelengths.

\section{ICS of self produced SR}

\subsection{Self produced SR}

When a CB encounters the wind/ejecta which was blown by the progenitor 
star long before the GRB, it sweeps in the ionized matter in front of it. 
The swept in electrons and nuclei that in its rest frame enter it with a 
Lorentz factor $\gamma$ equal to that of the bulk motion of the CB, are 
isotropized and Fermi accelerated in the CBs by its turbulent magnetic 
field. They emit SR with an early-time lightcurve
in the observer frame (DDD2009a) :
\begin{equation}
F_\nu[t] \propto  {e^{-a/t}\,
t^{1-\beta} \over t^2+t_{exp}^2}\, \nu^{-\beta}\!\rightarrow
        \! t^{\!-\!(1\!+\!\beta)}\, \nu^{\!-\!\beta}\, ,
 \label{SROP}
\end{equation}
where $t\!=\!T\!-\!T_i$, $T$ is the observer time 
after trigger and $T_i$ is the observer time when the CB reaches the 
wind/ejecta. The optical band  is initially  
well 
below the `bend' frequency (DDD2009a). Consequently, $\beta_O\!\approx\! 
0.5$ and the `prompt' optical flare that follows an ICS keV-MeV 
pulse/flare, decays like $F_\nu \! \propto\! 
t^{\!-\!1.5}\,\nu^{\!-\!0.5}\,.$

\subsection{ICS in the Thomson and Klein-Nishina regimes}

The differential cross section for Compton scattering
of a photon with energy $E_\gamma$  from an electron at rest 
was calculated  by Klein \& Nishina
in 1929. In the electron's rest frame it is given by,
\begin{equation}
{d\sigma \over d\Omega}={1 \over 2}\, \left ({e^2\over m_e\, c^2}\right)^2
\, (k-k^2\, sin^2\theta+k^3)\,,
\label{KN}
\end{equation}
where
\begin{equation}
k(E_\gamma,\theta)={1\over 1+{E_\gamma \over m_e\, c^2}\,(1-cos\theta)}
\label{KNk}
\end{equation}
The final energy of the scattered photon is, $k\,E_\gamma\, .$

In the Thomson regime where $E_\gamma\!\ll\! m_e\, c^2$,
and then  $k\!\approx\! 1$,
$d\sigma/ d\Omega\!\propto\!(1\!+\!cos^2\theta)$ and 
the mean final energy of a photon that suffered an ICS by an electron 
of energy $E_e$ is $(4/3)\, (E_e/m_e\, c^2)^2\, E_\gamma$.
A power-law distribution of Fermi accelerated electrons  $dn_e/dE\!
\propto\! E^{-p}$ generates through ICS a power-law  
distribution of scattered photons with $E^2\, dn_\gamma/dE\!\propto 
E^{-(p\!-\!3)/2},$  where $p\!\approx \!2.2$ well below 
the electron cooling energy and $p\!\approx \!3.2$ well above it.

In the Klein-Nishina (KN) regime where $E_\gamma\!>\! m_e\, c^2$
in the electron rest frame, 
$k\, E_\gamma \!\sim\!m_e\,c^2$,
$\sigma_{KN}\!\propto\!1/E_e$,
and the mean energy of a photon that suffered an ICS by an HE 
electron is approximately $E_e$. In the KN regime, 
a power-law distribution of electrons  $dn_e\!
\propto\! E^{\!-\!(p\!+\!1)}\, dE$ generates by ICS of self produced
SR a power-law
distribution of scattered photons with $E^2\, dn_\gamma/dE\!\propto
E^{\!-\!(p\!+\!1)/2)}$. In particular,  for a distribution of Fermi 
accelerated elecctrons 
that suffer a fast radiative cooling, $p\approx 3.2$ and
the scattered photons in the lab frame have a power-law distribution,
\begin{equation}
E^2\,{ dn_\gamma\over dE}\propto E^2\, {dn_e\over dE}\, 
\sigma_{KN} \sim E^{-(p+1)/2}\!\sim\! E^{-1.6}\, .
\label{KNFNU}
\end{equation}

\subsection{SSC in GRBs}

Like in blazars, the radiation produced by ICS of self produced SR,  the 
so called `SSC 
emission', is Doppler boosted by the CB relativistic motion and extends 
almost to 
TeV energies above which it is suppressed by the Klein-Nishina effect.
Roughly, the SSC energy flux 
$\nu\, F_\nu$ in the Thomson regime 
first increases with energy 
like $E^{(3\!-\!p)/2},$  where $p\!\approx \!2.2$ until 
$E\!\sim\! E_p\!\approx\!2\, m_e\,c^2\, \gamma\, \delta/3\,(1+z)$, then 
it changes into a plateau/shallow decrease 
like $E^{\!-\!(p\!-\!2)/2}$, due to the cooling break in the 
HE electron 
distribution,  until it enters the Klein-Nishina regime where it 
decreases like $E^{\!-\!(p\!+\!1)/2}.$
For typical GRBs  (DDD2009a) where  
$\delta\!\approx\!\gamma\!\sim\! 1000$ and $1\!+\!z\!\sim\!3.2$, the peak 
of the unabsorbed energy flux density is around $E_p\!\sim\! 100$ GeV. 
In very luminous GRBs, the peak energy $E_p$ of the SSC may
approach TeV, which is above the LAT energy range,  
and then could be detected only
in relatively very nearby GRBs where absorption in the IGM by  
pair-production 
can be neglected. Beyond the peak-energy, the SSC is cut off at an
energy $E_c\!=\! m_e\, c^2\, \gamma_{e,max}\, \gamma/(1\!+\!z)$
by the cut-off in the electron spectrum at 
$\gamma_{e,max}=\sqrt{6\,e/\sigma_T\,B_{eq}}$ in the CB rest frame  when 
the energy
loss-rate by SR exceeds the rate of Fermi acceleration by the strong
equipartition magnetic field $B_{eq}\!\approx\! \sqrt{4\,\pi\, n\,m_p\, 
c^2}\,\gamma$  generated in the CB by its collision
with the wind/ejecta ($n\, m_p $ is the density of the wind and
$\sigma_T\,\approx\! 0.665\times 10^{-24}\, {\rm cm^{-2}}$ is the 
Thomson cross section).

In the CB model each GRB pulse have a sub-MeV  energy flux density which 
is well 
approximated by an exponentially cutoff power-law
$F_E\!\propto\! E^{\!-\!\beta_g}\, e^{-E/E_p(t)}$ where 
$\beta_g\!\sim\! 0$, $E_p(t)\!\sim E_p(0)\,t_p^2/(t^2\!+\!t_p^2)$, 
and $t$ is the time after the beginning of 
the pulse (e.g. DDD2009a and references therein). Towards the end of 
each ICS pulse  simultaneous `prompt' optical and HE  
emissions begin. They 
have a power-law spectrum $F_{\nu}\!\propto\!\nu^{-\beta}$, with 
$\beta\!\sim\!0.5$ in the optical band and $\beta_O\leq \beta\leq \beta_X$ 
in the LAT band. 
The time integrated spectrum over a single GRB pulse (or several 
unresolved GRB 
pulses) appears to be an exponentially cutoff
power-law spectrum  with $E_p$ in the keV-MeV range  plus a 
power-law component. The power-law component dominates both at low 
energies 
because $\beta_g\lsim 0$ whereas  $\beta_{OX}\!\sim\! 0.8 $,
and at high energies because of the exponential cut-off of the spectrum 
of the prompt MeV emission.

\section{External ICS of prompt GRB photons}  

The thermal electrons (and the Fermi accelerated ones) in the CBs that 
have the CB bulk motion Lorentz factor $\gamma$ in the observer frame can 
boost by ICS the 
energy of prompt sub-MeV GRB photons which suffer Compton scattering in 
the wind 
and are overtaken shortly by the CBs. However, this mechanism is strongly 
suppressed by the Klein-Nishina effect for GRB  photons 
whose energy is above $\!\sim\!1$ eV and thus does not 
contribute effectively to the production of HE photons.

\section{Hadronic production of HE photons by thermal protons}
 
The radiative decay of $\pi^0$'s produced in  collisions 
between thermal CB nuclei and wind nuclei produce sub-TeV  $\gamma$-rays. 
Neglecting magnetic deflections of the CB nuclei in the wind,
the equivalent isotropic energy of sub-TeV
$\gamma$'s produced that way is given by,
\begin{equation}
E^{HE}_{iso}\sim 0.08\, E_{CB}\, \sigma_{in}\, N_{wind}\, \delta^2\,,  
\label{EISOMIN}
\end{equation}
where 0.08 is the fraction of the incident kinetic energy of protons that 
is carried by the $\pi^0$'s produced in HE pp collisions,  
$\sigma_{in}\!\approx\!40$ mbarn is the HE pp inelastic cross section,  
$N_{wind}\!\sim \! 10^{21\pm 1}{\rm cm^{-2}}$
is the typical column density of the GRB environment as inferred 
from the spectral measurement with the Swift X-ray telescope (XRT),
$E_{CB}\!\sim 10^{50}$ erg is the canonical kinetic energy of a CB ejected 
in core collapse supernovae which produce GRBs (DD2004), 
and $\delta^2$ is the relativistic beaming factor in the CB model. 
These
numbers yield $E^{\pi^0}_{iso}\!\sim\ 3.2\times 10^{50\pm 1}$ erg,
which is much smaller than the typically observed equivalent isotropic 
gamma ray energy of the HE component in GRBs with a detectable HE 
emission.

\section{Hadronic production of HE photons by Fermi accelerated protons}

The wind nuclei which enter a CB are Fermi 
acceleration through magnetic deflections
by the turbulent magnetic fields in the CB. Their total 
path-length inside a CB before they escape can be much larger than 
their mean free path for hadronic collisions. In that case, a fraction 
$\!\sim\! 0.08$ of the CB's kinetic energy is converted to HE gamma rays 
through $\pi^0$ production and decay and a similar fraction is converted 
through gamma ray production in the hadronic showers by radiative decays, 
SR from charged leptons and ICS of the SR in the CBs .  Due to Feynman 
scaling, these HE $\gamma$-rays have a power-law spectrum with a power-law 
index equal to that of the Fermi accelerated nuclei, i.e., 
$dn_\gamma/dE\!\propto\! E^{-2.2}$. The total equivalent isotropic energy 
is then bounded by a fraction $0.08\lsim k \lsim 0.25$ of the kinetic 
energy of the jet which is converted into HE $\gamma$-rays that are beamed 
into a solid angle $\sim \pi\, \delta^{-2}$. This yields the upper limit:  
\begin{equation} 
E^{HE}_{iso}\leq k\, \times 10^{56}\, \left[{E_{CB}\over 10^{50}\, 
   {\rm erg}}\right]\, \left[{\delta\over 1000}\right]^2\, {\rm erg}.  
\label{EISOMAX} 
\end{equation} 
In the CB model, the exact value of $k$ and consequently of $E^{HE}_{iso}$ 
and the early-time lightcurve of the HE `hadronic' $\gamma$ ray emission 
before the CBs enter the ISM are strongly model-dependent and involve many 
unknown parameters. However, when the jet enters the ISM, the lightcurve 
is similar to that of the X-ray afterglow until the jet break/bend 
(DDD2009a).

Roughly, the same spectra, $E_{iso}$, and lightcurves are expected for the 
HE $\gamma$-ray emission and the $\nu_\mu$ emission in GRBs from hadronic 
production and decay of $\pi^0$ and $\pi^{\pm}$, respectively.

\section{Comparison with observations}

\begin{figure}[]
\centering
\epsfig{file=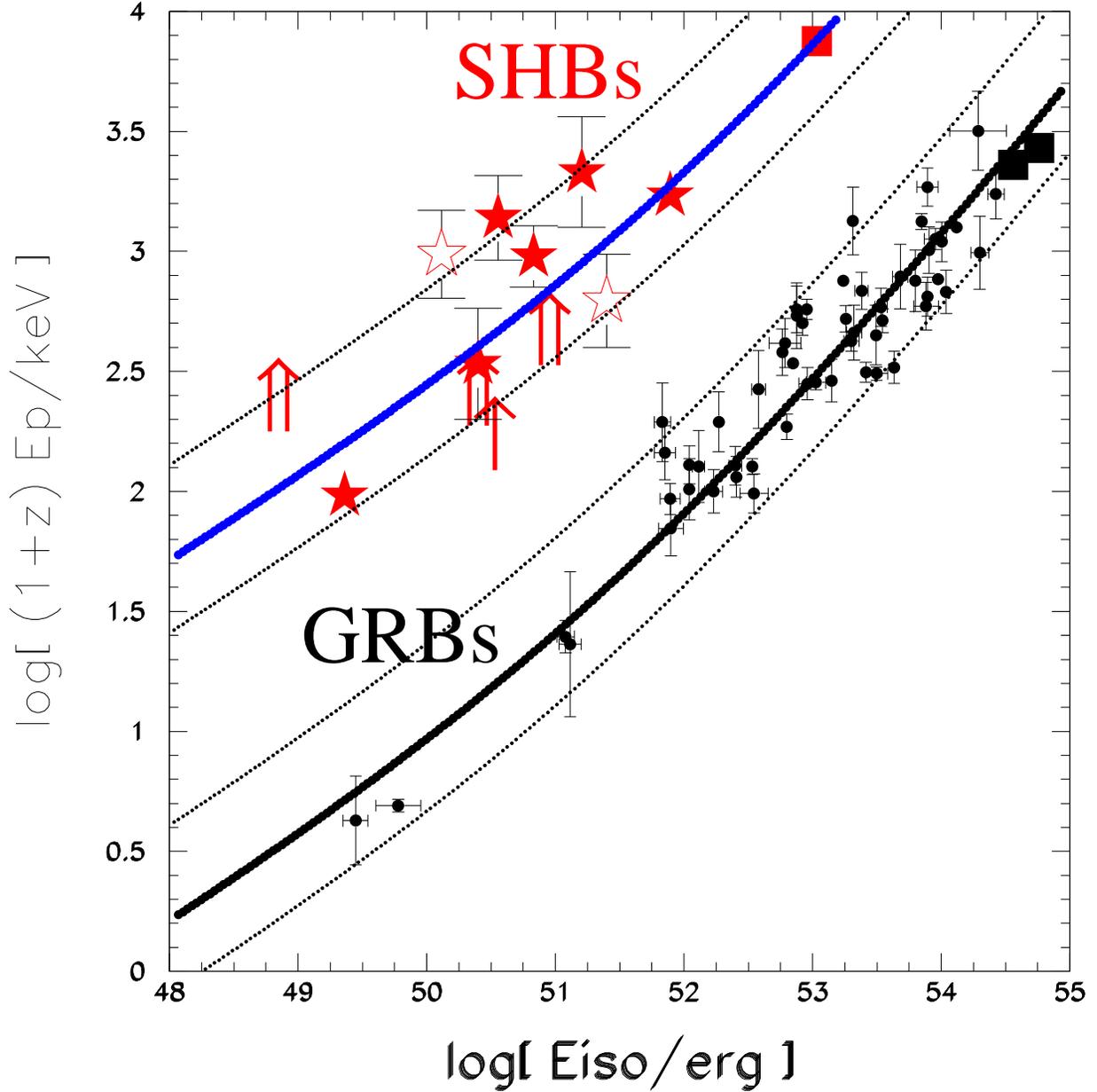,width=18cm}
\caption{
Comparison between the observed correlation [Ep,Eiso] in LGRBs and SHBs 
and the predicted correlation for LGRBs (DDD2007, Eq.(4)) and SHBs  
(DDD2009b, Eq.(22)). GRB 080916C, GRB 090902B and SHB 090510
are indicated by full squares (black, black and red, respectively, in the 
color version).
}
\label{f1}
\end{figure}

The rapid  localization of a few degrees,
which is delivered by the Fermi GBM, is not precise enough for 
a rapid localization and follow-up by current optical telescopes. 
The localization of the HE emission detected by the 
LAT seems
good for optical follow-up  but, so far, it is delivered with a 
large delay compared to the GRB
itself. Consequently, the coincidence in time between the HE emission
and the optical emission and the similarity between their   
light curves, which are predicted by the CB model, cannot be tested
yet in individual GRBs.  However, the decay of the prompt 
optical flares seems to satisfy the CB model prediction (DDD2009)
of a power-law decay with an index $\alpha$=1+$\beta_0\!\sim 1.5$.
Such a power-law decay  seems to agree with the observed power-law decay 
of the HE emission 
that was detected by the Fermi LAT in 11  GRBs so far
(see, e.g., Ghisellini1, Ghirlanda \& Nava 2009).   
Moreover, the spectrum of the HE emission which was measured by the Fermi 
LAT  seems to be described well by 
a power-law with a spectral index which satisfies, admittedly 
within large errors, the relation, 
$\Gamma_{LAT}\!\approx\!\Gamma_X$, predicteby the CB model,
where $\Gamma_X$ is the photon index of the {\it late-time X-ray AG}.
This is shown in Table~\ref{t2} and in Fig.~\ref{f1} where
we plotted the ratio $\Gamma_{LAT}/\Gamma_X$ for  7 GRBs 
that were observed both by the Fermi LAT and by the Swift XRT.  
Note that the average ratio  is consistent with 1 with a small error
unlike large error in this ratio for the individual GRBs.
Below we compare in detail the CB model predictions 
and the observations of two representative GRBs with HE emission
that was detected by LAT, the long GRB 090902B and SHB 090510.

\subsection{GRB 090902B}

\noindent
{\bf Observations:} The Fermi Gamma-ray Burst Monitor (GBM) triggered on 
and localized the bright burst GRB 090202B on 2009 September 2 at 
11:05:08.31 UT. The burst had a multi-peak structure with the brightest 
peak around 14 sec after trigger. It was also detected by the INTEGRAL and 
Suzaku satellites.  The sub-MeV ended approximately 25 seconds after 
trigger. Emission above 100 MeV was detected by the Fermi LAT up to 1k 
sec after the GBM trigger when the Earth's limb was starting to enter its 
field-of-view, with 39 photons above 1 GeV (de Palma et al.~2009). The 
highest energy photon had $E$= 33.4 +2.7/-3.5 GeV and arrived 82 seconds 
after the GBM trigger. The emission declined like $t^{\!-\!1.5}$ until the 
LAT observations were interrupted by entry of the Earth's limb into its 
field of view, but analysis of data an hour after trigger, when the source 
location was again unocculted, showed that any later emission lied below 
the LAT sensitivity (Bissaldi et al.~2009). At a redshift of $z$=1.822 
(Cucchiara et al.~2009), the fluence of $(4.36\pm 0.06)\times 10^{-4}\, 
{\rm erg\, cm^{-2}}$ during the first 25 seconds of the prompt emission 
yields  $E_{iso}\!=\!(3.63 \pm 0.05)\times 10^{54}$ erg
isotropic equivalent $\gamma$-ray energy in the 10 keV-10 GeV 
range.

The X-ray follow-up observations with the Swift X-ray telescope 
(XRT)  started only 12.5 hours after the GBM trigger. The measured X-ray 
spectrum can be fit by an absorbed power-law model with a photon index of 
$\Gamma\!=\! 2.1\pm 0.3$ and a rest frame column density of $(3.4 \pm 
0.9)\times 10^{22}\, {\rm cm^{-2}}$ at $z$=1.822 in addition to the 
Galactic column density in the direction of the burst (Stratta et al.~2009). 

The earliest ground-based optical observations were obtained 
only $\!\sim\!1.4$ h after trigger by ROTSE-IIIa 
(Pandey et al.~2009). The burst was later 
detected also in the optical, infra red and radio.

\noindent
{\bf Interpretation:} From the LAT detection of
the highest energy photon of 11.16 (+1.48/-0.58) GeV  
during the prompt phase  Bissaldi et al.~(2009) inferred a  
minimum value of the bulk Lorentz factor $\gamma(0)\! \geq\! 1000$
of the source using the flux variability time scale of 53 ms
and the constraint that the opacity for $e^{\pm}$ pair-production    
in the source is less than unity so that such photons can escape outside 
the source (Fenimore et al.~1993). A similar value was
derived for GRB 080916C (Abdo et al.~2009a).
{\it Such values of the bulk motion Lorentz 
factor of the jet of CBs were long advocated by the CB model
(e.g., DDD2002; DD2004) and were used to explain the 
typical photon energy and isotropic equivalent energy of long GRBs.} 
For standard candle GRBs, the largest observed values of $E_{iso}$ and 
$E_p$ are obtained when the GRBs are viewed from very near axis, i.e., 
with a viewing angle $\theta^2\!\ll\! 1/\gamma(0)^2$.
In that case the Doppler factor of the CBs is $\delta(0)\!\approx \! 
2\, \gamma(0)$. In the CB model, the isotropic gamma ray energy
of such GRBs is (DD2004): 
$E_{iso}\!\approx\!  8\times 10^{52}\,(\delta(0)/1000)^3\, N_{CB}$ 
erg. It  yields $E_{iso}\!\geq\! 3.2 \times 10^{54}$ erg
for  a jet of $N_{CB}$=5 canonical CBs (producing the 5 
main peaks in the lightcurve of GRB 090902B) 
and $\delta(0)\!=\!2\, \gamma(0)\!\geq\! 2000$. This value agrees 
with the value  $E_{iso}\!\approx\!3.63\times 10^{54}$ erg that was  
inferred by Bissaldi et al.~(2009) from the Fermi measurements.

\begin{figure}[]
\centering
\epsfig{file=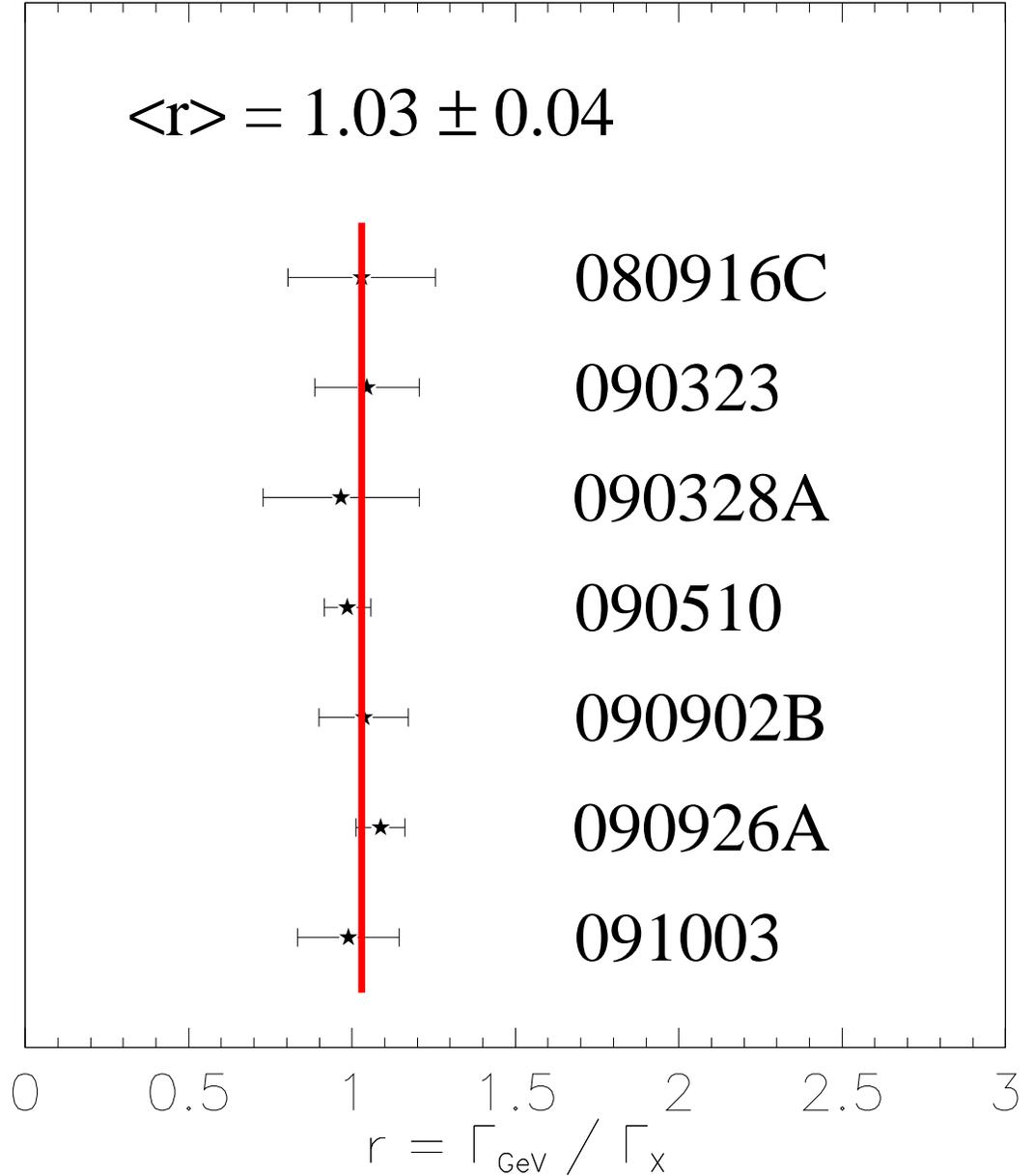,width=18cm}
\caption{
The ratio of the photon spectral index of the prompt HE emission 
detected by the Fermi LAT in several GRBs  and the spectral index
(Ghisellini et al.~2009)
of their unabsorbed late-time X-ray AG inferred from  the Swift XRT
measurements and reported in the Swift Lightcurve Repository,
Evans et al.~2009. 
}
\label{f2}
\end{figure}

The time-averaged spectrum in the 100-1000 keV during the 25 sec
of the sub-MeV GRB prompt phase was well 
fitted by a power-law with photon index $\Gamma$=0.91+/-0.10 
and an exponential cutoff with $E_p$=885+/-0.39 keV
(Terada~et al.~2009). The photon index 
$\beta_g$=$\Gamma$-1=-0.09+/-0.10 
is consistent with the CB model expectation $\beta_g\!\sim\!0$.
The rest frame  value of the peak energy 
$(1\!+\!z)\,E_p\!=\!2550 \pm 112$ keV 
is also consistent with that expected in the CB model
as shown in Fig.~(\ref{f2}):  it satisfies 
the so called `Amati correlation' which  
follows from the CB model and was predicted
(see e.g.,  DDD2007
and references therein) long before it was discovered. 

The GRB peaks are not well resolved in the Fermi GBM data and do not allow 
a stringent test of the CB model predictions for the temporal and 
spectral behaviour of the individual prompt MeV peaks/flares.

\begin{figure}[]
\centering
\epsfig{file=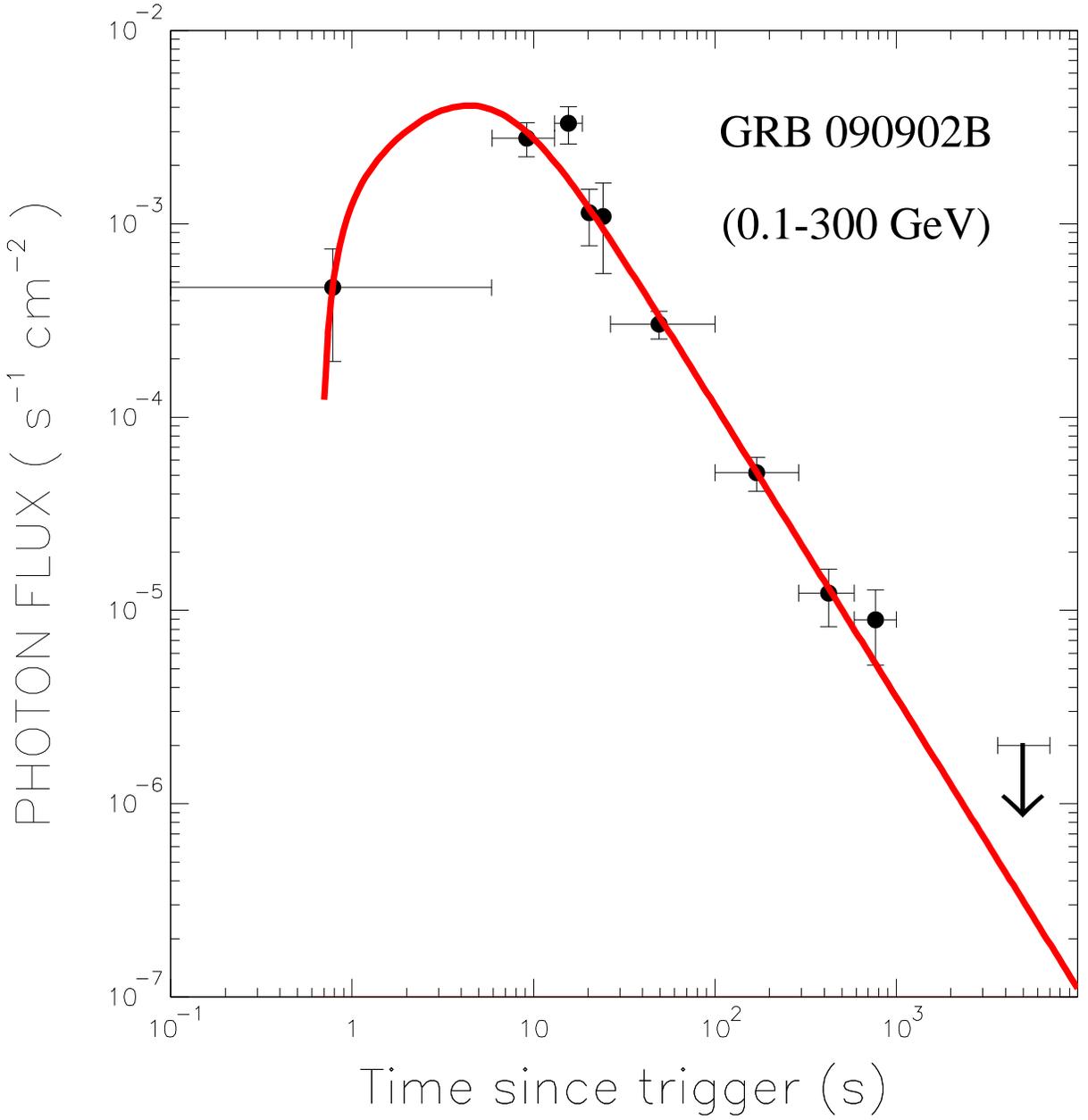,width=18cm}
\caption{Comparison between the FERMI LAT 
lightcurve of the high energy emission in 
GRB 090902B (Bissaldi et al.~2009 )
and its CB model description, Eq.~(\ref{ICSOP}). 
}
\label{f3}
\end{figure}

Due to insufficient statistics, also the early-time lightcurve
of the HE emission, which was 
measured by the Fermi LAT, was not well resolved into separate peaks.  
However, as shown in Fig.~(\ref{f3}), it is well described by 
Eq.~(\ref{ICSOP}) 
assuming  a single effective CB with the parameters listed in 
Table~\ref{t3}.
Note that its predicted decline, until taken over by its shallow 
decay of the SSC in the ISM is well described by the expected decline, 
$F_\nu\!\propto\! t^{\!-\!1!\-\!\beta_O}$ (see Eq.~(\ref{ICSOP}).  
Also the photon spectral index of the 
HE emission as measured by Bissaldi et al.~2009, 
$\Gamma\!=\!1.93\pm 0.03$, as predicted, 
is equal within errors to the measured spectral index in the X-ray band,
$\Gamma_X\!=\!2.1\pm 0.3$ inferred by
Stratta et al.~(2009) from the Swift XRT data.

No early-time optical data is available to test the CB model prediction 
that the prompt optical emission and the high energy emission have the 
same temporal behaviour. In the CB model each prompt MeV peak/flare has a 
delayed HE peak that decays asymptotically like $t^{\!-\!1\!-\!\beta_O}$ 
and the time-lag between the prompt MeV emission and the high energy 
emission in each pulse is of the order of the pulse width, while the bulk 
of the high energy emission is delayed typically by a time comparable to 
the duration of the MeV component. There are indications for such 
correlations in the data obtained by the Fermi LAT and GBM on GRB 090902B 
and 080916C.

\begin{figure}[]
\centering
\epsfig{file=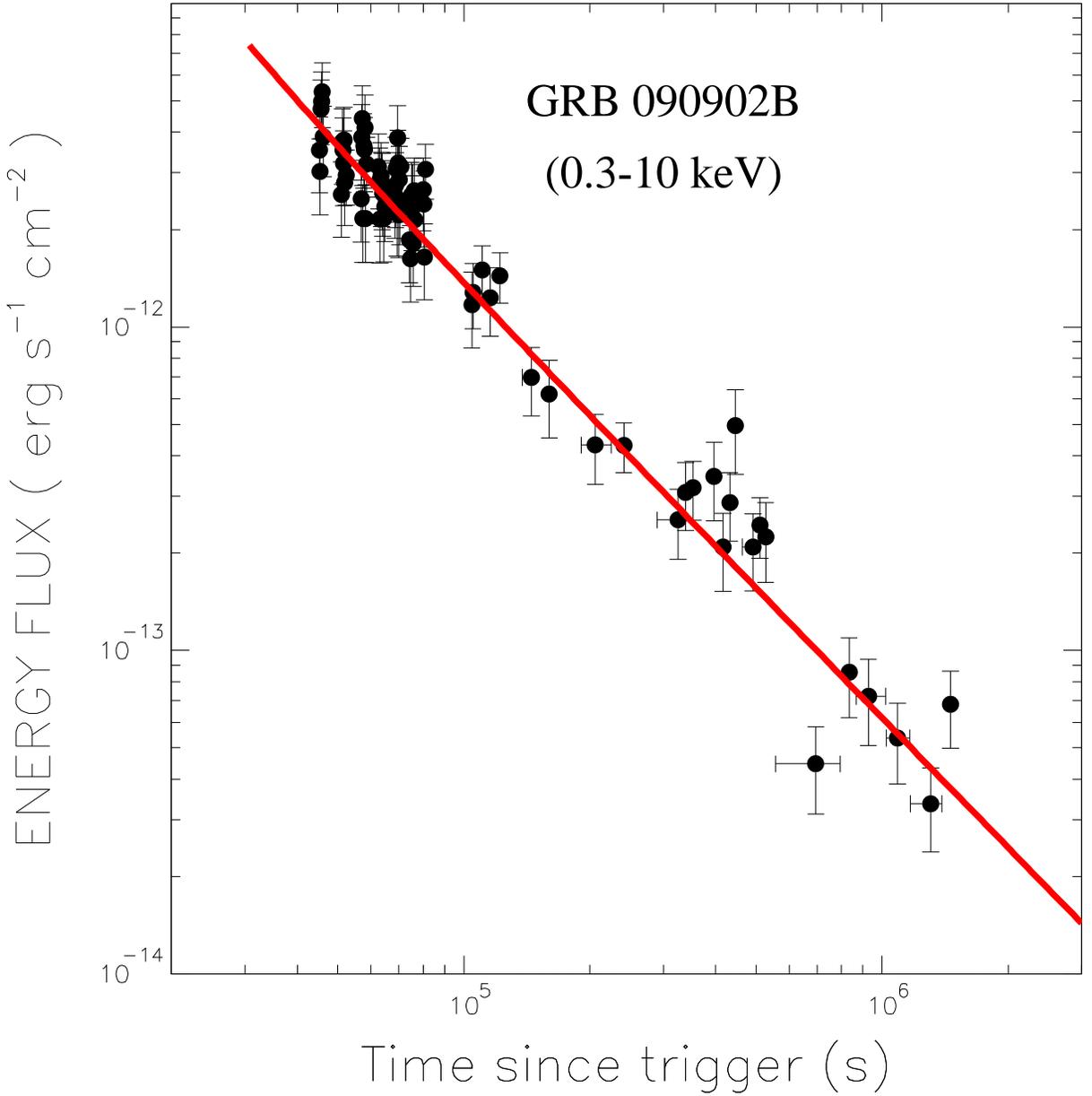,width=18cm}
\caption{Comparison between the Swift XRT lightcurve of
GRB 090902B (Evans et al.~2009)
and its CB model description
The asymptotic power-law decline with $\alpha_X\!=\!1.42\pm 0.10 $ and 
$\beta_X\!=\!1.10\pm0.30 $ satisfies 
the CB model prediction $\alpha_x \!=\!\beta_X\!+\!1/2$ for a constant 
density ISM.
}
\label{f4}
\end{figure}

In the CB model, the observed late time behaviour of the lightcurve of the 
unabsorbed X-ray afterglow is described by a power-law $F_\nu\!\propto\! 
t^{\!-\!\alpha_X}\,\nu^{\!-\!\beta_X}$ with $\alpha_X$=$\beta_X$+1/2.  A 
best fit to the Swift XRT data shown in Fig.~(\ref{f4}) yields 
$\alpha_X\!=\!1.42\pm 0.1$. This value is consistent within errors with 
$\alpha_X\!=\!1.6\pm 0.3$ expected from the spectral index 
$\beta_X\!=\!1.1\pm 0.3$ inferred by Stratta et al.~(2009) from the Swift 
XRT data.

\subsection{SHB 090510}

{\bf Observations:}
SHB 090510 is a short/hard burst at redshift
$z$=0.903 (Rau et al.~2009) detected by Fermi
(Guiriec et al.~2009), AGILE (Longo et al.~2009), Swift
(Hoversten et al.~2009), Konus-Wind (Golenetskii et al.
2009) and Suzaku (Ohmori et al.~2009). The Fermi GBM triggered
on a precursor while the main emission episode 
started $\sim 0.5$ sec after trigger and lasted $\sim 0.2$ sec.
Its lightcurve consists of 7 main peaks.
The emission detected by the Fermi LAT started 0.65 s after the trigger
and lasted $\sim 200$ sec.
The photon with the highest energy  
of $31 \pm 3$ GeV arrived 0.829 sec 
after trigger (Abdo et al.~2009b).
The equivalent isotropic gamma ray energy before the onset of the 
high energy emission was $E_{iso}$=(3.91 -0.88/+1.91)$\times 10^{52}$ erg.
The time integrated photon spectrum measured by AGILE
(Giuliani et al.~2009) was fitted with an exponentially 
cutoff power-law with a photon power-law index $\Gamma$=0.65 (-0.32/+0.28)
and a cutoff energy $E_c$=2.8 (-0.6/+0.9) MeV 
(rest frame peak energy $E_p$=7.19 (-1.54/+2.31) MeV).
The time integrated spectrum between 0.5 and 1 sec after trigger
was best fitted by adding a second component -a power-law component with 
a photon index $\Gamma\!=\!1.62 \pm 0.03$- and resulted with 
$E_p\!=\!3.9 \pm 0.3$ MeV and $E_{iso}\!=\!(1.08 \pm 0.06)\times 10^{53}$
erg (Abdo et al~2009b). The high-energy spectral component accounts
for $\sim 37\%$ of the total fluence.
The bulk of the photons above 30
MeV arrived $258 \pm 34$ ms later than those below 1 MeV.
 
{\bf Interpretation:}

The detection of a 31 GeV photon during the first second sets the highest 
lower limit on the bulk motion Lorentz factor of the source, 
$\gamma\!>\!1200$ (Abdo et al.~2009b). This limit is consistent with the 
typical value $\gamma(0)\!\sim\!1400$ advocated by the CB model for 
ordinary SHBs (DDD2009b). The large value of $E_p$ probably requires 
$\gamma(0)\!\sim\!2000$ and a small viewing angle $\theta^2\ll 
1/[\gamma(0)]^2$ implying $\delta(0)\!\approx \! 2\, \gamma(0)$ and 
$E_{iso}$ larger roughly by a factor 23 than $5\times 10^{51}$ erg, the 
mean $E_{iso}$ of SHBs.

Due to insufficient statistics, the early-time lightcurve
which was measured by the Fermi LAT cannot 
be resolved reliably into separate HE peaks. The 
lightcurve of the blended peaks inferred from the LAT measurements
is well described by Eq.~(\ref{ICSOP}) for a single effective CB with the
parameters listed in Table~\ref{t3} as shown in Fig.~(\ref{5}).  
Also the photon spectral index of the 
HE emission as measured by Giuliani et al.~(2009), 
$\Gamma\!=\!1.58$ (-0.11,+0.13), and by
Bissaldi et al.~(2009), 
$\Gamma\!=\!1.62\pm 0.03$, 
is roughly within errors that of the spectral index 
$\Gamma_X\!=\!1.792$ (+0.071/-0.051) 
inferred by Evans et al.~(2009) from the Swift XRT data.
No early-time optical
data is available to test the CB model prediction that the prompt optical
and high energy emissions have the same temporal behaviour.
The two spectral components (a prompt MeV cutoff power-law component and 
a delayed HE power-law component), 
show  a significant temporal correlation (Abdo et al.~2009b)
as expected in the CB model.
Lag-times within the MeV band and the HE band 
were not detected, as expected in the CB model (DDD2009b).
The bulk of the photons above 30
MeV arrived $258 \pm\ 34$ ms later than those below 1 MeV. This 
time lag is comparable to the duration of the sub-MeV emission,
as expected in the CB model. 

In the CB model the asymptotic decline of the X-ray afterglow in  
$n\!\propto\! 1/(r\!-\!r_c)^2$ density beyond $r_c$ is given by
$F_\nu\!\propto\! t^{\!-\!\alpha}\,\nu^{\!-\!\beta}$
with $\alpha$=$\beta$+1=$\Gamma$ (e.g., Eq. 32 in DDD2009b).
The Swift XRT lightcurve repository  (Evans et al.~2009)
reports $\Gamma$=1.792(+0.071/-0.051) from a spectral analysis of the 
X-ray afterglow, in good agreement with the index 
$\alpha$=1.89$\!\pm\!$0.06 
of the best fit power-law decline beyond 1100 sec after trigger
(see Fig.~(\ref{6})).

\section{Conclusions}

In the cannonball (CB) model,
high energy emission from GRBs and SHBs is a natural consequence of the   
model. The dominant leptonic and hadronic emission mechanisms are ICS  
of SR by Fermi accelerated
electrons in the collision of the jet of highly relativistic CBs with
the wind/ejecta blown from the progenitor or companion star long before
the GRB, and the decay of $\pi^0$ produced in hadronic collisions
between the CB nuclei and the nuclei of the hadronic matter
(ejecta/wind/ISM) that the jet passes through. The main predictions of the
model  for the early-time high energy emission are:   
\begin{itemize}
\item{}
Each prompt keV-MeV  pulse is followed by a delayed HE emission
which lasts much longer.    
  
\item{}
The HE emission coincides in time with the optical emission.

\item{}
The light curve of the HE emission is roughly proportional to that of 
the unextinct optical emission. The decay of both the `prompt' optical
and the HE emissions is a power-law with an index
$\alpha$=1+$\beta_0\!\sim 1.5$. 
 
\item{} 
The spectrum of the HE component is a simple power-law 
with a spectral index approximately equal to that of the X-ray afterglow, 
i.e., $\Gamma_{LAT}\!\approx\!\Gamma_{X}$.

\item{} 
The HE emission extends to very high energies,
where the observed
radiation is strongly attenuated by $e^+e^-$ pair productin in the EBL.

\item{}
The equivalent isotropic energy of the HE component in bright GRBs can 
reach, and even exceed that in the sub-MeV range.

\item{}
The neutrino counterpart of the hadronic emission of HE $\gamma$-rays 
from the most luminous GRBs is barely detectable in $km^3$ 
underwater/under-ice Cherenkov neutrino telescopes (DD2008).

\end{itemize}
\noindent
These predictions are consistent with the present available
data on high energy emission from GRBs obtained from the
gamma ray satellites and from the large air, ground and underground
Cherenkov telescopes. In particular, the main observed properties
of the HE emission measured with the Compton, Fermi and AGILE gamma ray
satellites are well reproduced by the CB model as demonstrated
in this paper for GRB 090902B and SHB 090510.

An observational proof of the ICS origin of the HE gamma ray emission from 
GRBs requires simultaneous detections of the prompt optical and HE 
emissions. It is highly desireable that the LAT team improves their 
automatic analysis, in order to deliver a GRB position within a few 
seconds after the LAT detection. Even a few tens of seconds will be very 
useful. This will provide a stringent test of models of HE emission from 
GRBs (and from other HE transient sources such as blazars, microquasars, 
pulsars, etc) and help pin down their production mechanism. Extremely 
optically-luminous GRBs, such as GRB 080319 where the prompt optical 
emission was resolved into individual flares (Racusin et al.~2008), may 
show that also the prompt high energy emission consists of HE flares which 
are associated with and follow promptly each individual keV-MeV pulse, as 
expected in the CB model.

\begin{deluxetable}{llllc}
\tablewidth{0pt}
\tablecaption{The dominant particle populations in the CBs and their
dominant radiation mechanisms during the prompt/early-time emission phase 
in GRBs}
\tablehead{
\colhead{CB Population} & \colhead{Origin} & \colhead{Target} & 
\colhead{Mechanism} & \colhead{Energy Band}
 }
\startdata
Thermal e's & CB  & Glory photons & ICS & keV-MeV \\
Fermi accelerated e's& CB-Wind Collision & CB Magnetic Field & SR & UVOIR \\
Fermi accelerated e's & CB-Wind Collision & Self SR  & ICS (SSC) & HE \\
Thermal p's        & CB & Wind/Ejecta & $\pi^0$ decay & Sub-TeV\\ 
Fermi accelerated p's &CB-Wind Collision & CB Nuclei   &  $\pi^0$ decay& UHE \\
\enddata
\label{t1}
\end{deluxetable}

\begin{deluxetable}{llllc}
\tablewidth{0pt}
\tablecaption{The dominant particle populations in the CBs and their
dominant radiation mechanisms during the prompt/early-time emission phase
in GRBs}
\tablehead{
\colhead{GRB} & \colhead{$T_i$ [s]} & \colhead{$T_f$ [s]} & 
\colhead{$\Gamma_{LAT}$} & \colhead{$\Gamma_X$} 
 }
\startdata
080825C  &   0  &  200 &  $1.96 \pm 0.30$ &      \\
080916C  &   0  &  200 &  $2.09 \pm 0.12$ & 2.03 (+0.43, -0.43) \\
081024B  &   0  &    5 &  $1.64 \pm 0.47$ &      \\
090217   &   0  &    1 &  $2.22 \pm 0.40$ &      \\
090323   &   0  &  400 &  $2.05 \pm 0.20$ & 1.96 (+0.24, -0.21) \\
090328   &   0  &  100 &  $1.76 \pm 0.35$ & 1.82 (+0.23, -0.31) \\
090510   &   0  &    7 &  $2.15 \pm 0.12$ & 1.79 (+0.071, -0.051) \\
090626   &   0  &  600 &  $1.70 \pm 0.12$ &      \\
090902B  &   0  &  320 &  $2.32 \pm 0.16$ & 2.24 (+0.25, -0.25) \\ 
090926A  &   0  &   25 &  $2.34 \pm 0.14$ & 2.15 (+0.072, -0.071) \\
091003   &   0  &  100 &  $1.85 \pm 0.25$ & 1.87 (+0.15, -0.16)   \\
\enddata
\label{t2}
\end{deluxetable}

\begin{deluxetable}{lllllllc}
\tablewidth{0pt}
\tablecaption{The parameters of the CB model fitted
lightcurve of the prompt/early-time HE emission and the late X-ray AG 
from GRB 090902B and SHB 090510.}
\tablehead{\colhead{GRB} & \colhead{$T_i$ [s]} & \colhead{$t_{exp}$ [s]} &
\colhead{a $[s]$} &  \colhead{A} &\colhead{$\beta_O$}& 
\colhead{$\beta_{HE}$} &\colhead{$\beta_X$} 
} 
\startdata
GRB 090902B & 0.63 & 6.28 & 0.14&  0.118 $cm^{-2}\, s^{-1}$ &
0.51 & $0.93\pm 0.03$ & $1.10\pm 0.30$ \\
\hline
SHB 090510 & 0.662 & 0.227 & 0.0167 & 36.4 $s^{-1}$ & 
0.54 & $0.62\pm 0.03$ & $0.79\pm 0.07$  \\
\enddata
\label{t3}
\end{deluxetable}

\newpage

\newpage

\newpage

\newpage

\newpage

\newpage
\begin{figure}[]
\centering
\epsfig{file=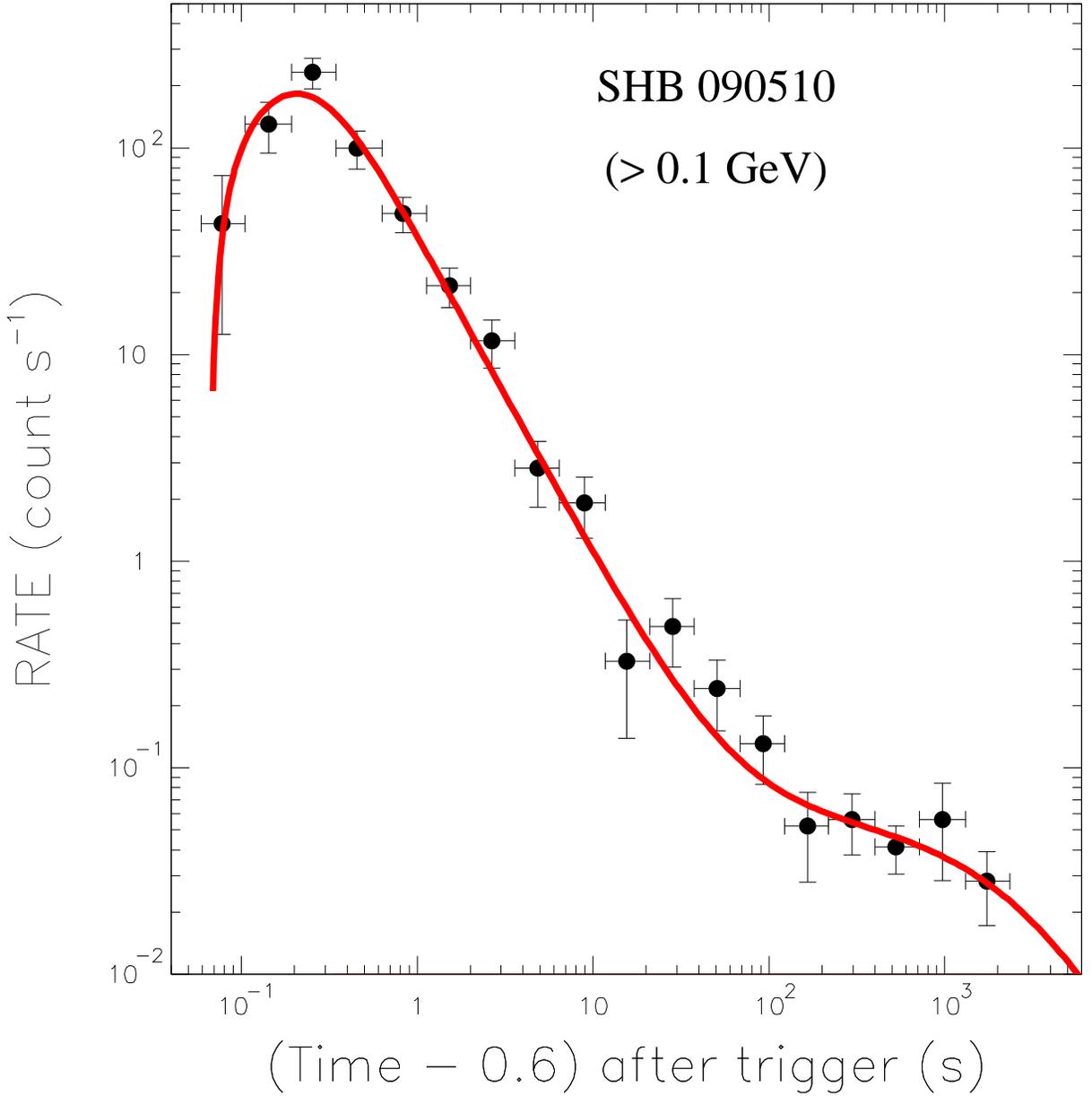,width=18cm}
\caption{Comparison between the FERMI LAT
lightcurve of the high energy emission in
SHB 090510 (Ghirlanda et al.~2009)
and its CB model description as given by Eq.~(\ref{ICSOP}).
}
\label{f5}
\end{figure}

\newpage
\begin{figure}[]
\centering
\epsfig{file=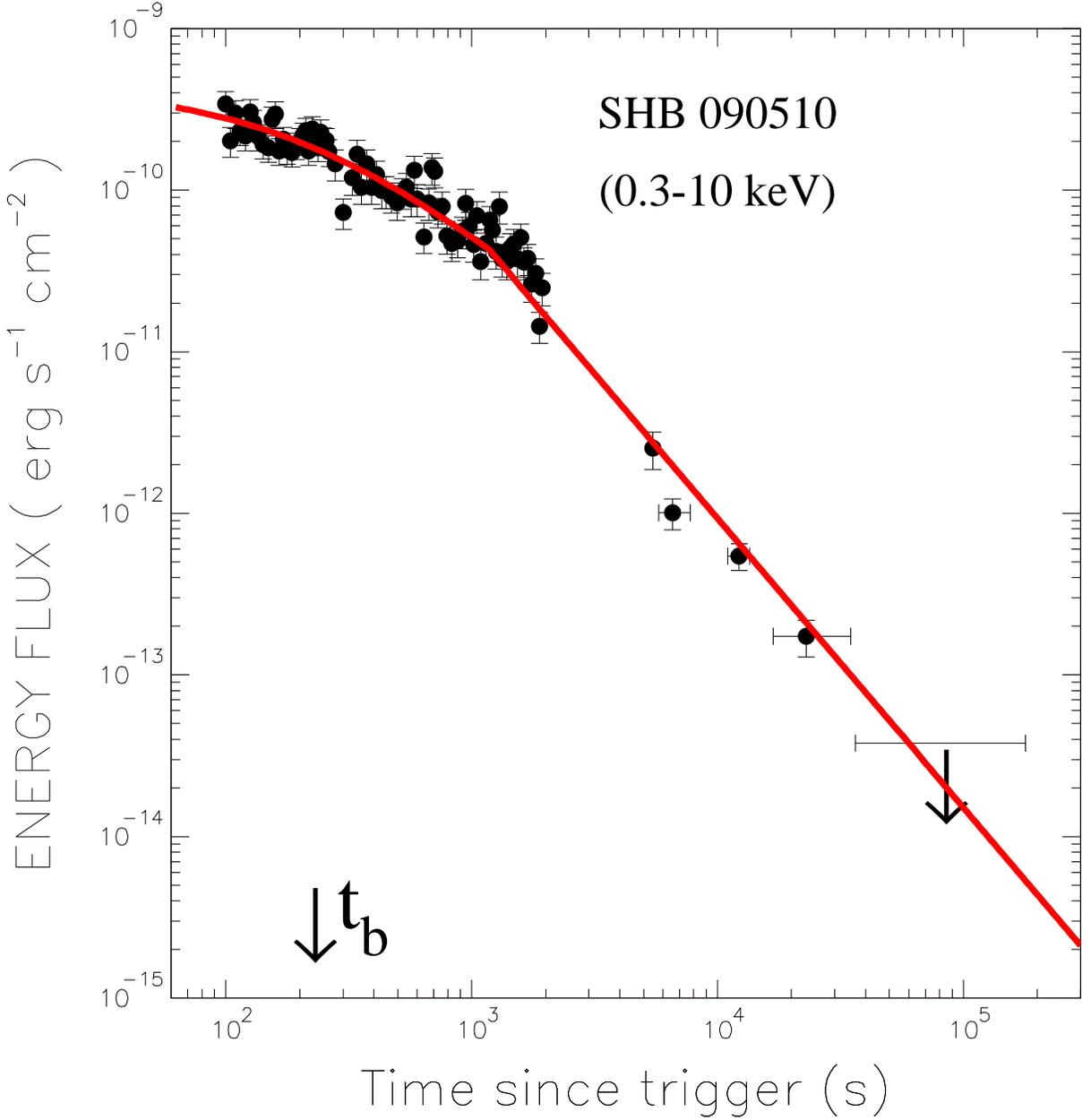,width=18cm}
\caption{Comparison between the Swift XRT lightcurve of SHB 090510
(Evans et al.~2009) and its CB model description.
The asymptotic power-law decline with $\alpha\!=\!1.89 \pm 0.06 $ and
$\beta\!=\!0.792$(+0.71/-0.051) satisfies within errors
the CB model prediction $\alpha \!=\!\beta\!+\!1\!=\!\Gamma $ for a 
density profile
$n\!\propto\! 1/(r\!-\!r_c)^2$ beyond $r_c$ (see DDD2009a,b). 
}
\label{f6}
\end{figure}

\end{document}